\documentclass[12pt,a4]{article}
\usepackage{sw20lart}

\input tcilatex
\begin{document}

\title{\textbf{A simple remark on three dimensional gauge theories }}
\author{\textbf{V.E.R. Lemes, C. Linhares de Jesus, C.A.G. Sasaki, } \and \textbf{%
S.P. Sorella, L.C.Q.Vilar} \\
UERJ, Universidade do Estado do Rio de Janeiro\\
Departamento de F\'\i sica Te\'orica\\
Instituto de F\'{\i}sica\\
Rua S\~ao Francisco Xavier, 524\\
20550-013, Maracan\~{a}, Rio de Janeiro \vspace{2mm}\\
and\vspace{2mm} \and \textbf{O.S.Ventura} \\
CBPF, Centro Brasileiro\textbf{\ }de Pesquisas F\'{\i}sicas \\
Rua Xavier Sigaud 150, 22290-180 Urca \\
Rio de Janeiro, Brazil\vspace{2mm} \and \textbf{CBPF-NF-050/97}\vspace{2mm}%
\newline \and \textbf{PACS: 11.10.Gh}}
\maketitle

\begin{abstract}
Classical three dimensional Yang-Mills is seen to be related to the
topological Chern-Simons term through a nonlinear but fully local and
covariant gauge field redefinition. A classical recursive cohomological
argument is provided.

\setcounter{page}{0}\thispagestyle{empty}
\end{abstract}

\vfill\newpage\ \makeatother
\renewcommand{\theequation}{\thesection.\arabic{equation}}

\section{\ Introduction\-}

Since many years topological three dimensional massive Yang-Mills theory 
\cite{tmym} is a continuous source of investigation and has led to a large
amount of interesting applications in different areas of theoretical physics.

As it is well known, the action of the model is characterized by two
parameters $(g,m)$, a gauge coupling and a mass term respectively, and can
be written as the sum of a Yang-Mills and of a Chern-Simons term, \textit{%
i.e.}

\begin{equation}
\mathcal{S}_{YM}(A)\;+\;\mathcal{S}_{CS}(A)\;,  \label{tmym}
\end{equation}
where, adopting the same parametrization of refs.\cite{gmrr,lm},

\begin{equation}
\mathcal{S}_{YM}(A)=\frac 1{4m}tr\int d^3xF_{\mu \nu }F^{\mu \nu }\;,
\label{ym}
\end{equation}
and

\begin{equation}
\mathcal{S}_{CS}(A)=\frac 12tr\int d^3x\varepsilon ^{\mu \nu \rho }\left(
A_\mu \partial _\nu A_\rho +\frac 23gA_\mu A_\nu A_\rho \right) \;,
\label{cs}
\end{equation}
with.

\begin{equation}
F_{\mu \nu }=\partial _\mu A_\nu -\partial _\nu A_\mu +g[A_\mu ,A_\nu ]\;.
\label{fs}
\end{equation}
Although being only power counting superrenormalizable, topological massive
Yang-Mills $\left( \text{\ref{tmym}}\right) $ turns out to be ultraviolet
finite to all orders of perturbation theory. This remarkable feature was
first detected by explicit one loop computations \cite{pr} and later on has
been put on firm basis and extended to all orders by \cite{gmrr} with a
careful study of the behaviour of higher loops three dimensional Feynman
integrals. More recently, this result has been proven to hold \cite{rrvn} by
similar arguments for the $N=1$ supersymmetric version of $\left( \text{\ref
{tmym}}\right) $.

It is also worthwhile to mention that a partial proof of the ultraviolet
finiteness of the $N=2$ version of $\left( \text{\ref{tmym}}\right) $ in the
Wess-Zumino gauge has been achieved in \cite{mpr} with a purely algebraic
cohomological analysis. The interesting result obtained here is that the
possible invariant counterterms turn out to be related to only one of the
two parameters $(g,m)$.

Let us come now to the main purpose of this work. Our aim here is to report
on a very elementary classical geometrical aspect which we shall hope to be
useful for a better understanding of the model. We shall be concerned, in
particular, with the observation that the topological massive Yang-Mills
action $\left( \text{\ref{tmym}}\right) $ can be actually traced back to a
pure Chern-Simons action through a nonlinear redefinition of the gauge
field, namely

\begin{equation}
\mathcal{S}_{YM}(A)\;+\;\mathcal{S}_{CS}(A)=\mathcal{S}_{CS}(\widehat{A})\;,
\label{mf}
\end{equation}
with

\begin{equation}
\widehat{A}_\mu =A_\mu +\sum_{n=1}^\infty \frac 1{m^n}\vartheta _\mu
^n(D,F)\;  \label{mf1}
\end{equation}
where the coefficients $\vartheta _\mu ^n(D,F)\;$turn out to be \textit{%
local }and \textit{covariant}, meaning that they are built up only with the
field strength $F$ and the covariant derivative $\left( D_\mu =\partial _\mu
+g[A_\mu ,\;]\right) $. The above formulas represent the essence of the
present letter. Their meaning is that, at the \textit{classical level},
topological massive three dimensional Yang-Mills can be seen as being
related, through a nonlinear but \textit{local} and \textit{covariant} field
redefinition, to the topological Chern-Simons term.

However, before going any further, some necessary remarks are in order. We
underline that the equation $\left( \text{\ref{mf}}\right) $ has to be
understood here in pure classical terms. Although the quantum aspects are
out of the aim of this work, let us observe here that at the level of the
quantized theory the nonlinear field redefinition $\left( \ref{mf1}\right) $
could seem to allow for a transfer of the properties of topological massive
Yang-Mills theory from the initial action $\left( \text{\ref{tmym}}\right) $
to the gauge fixing and the Faddeev-Popov terms. However, as one can easily
infer from the presence of the expansion parameter $1/m$ in the eq.$\left( 
\ref{mf1}\right) $, the use of the redefined gauge field $\widehat{A}_\mu $
will introduce in these terms an infinite number of power counting
nonrenormalizable interactions which would render the quantum analysis more
involved. In other words, as far as the quantum aspects are concerned, the
use of a manifest power counting renormalizable gauge coordinate system and
of the usual action $\left( \text{\ref{tmym}}\right) $ as the starting
points are more suitable, as proven by \cite{tmym, pr,gmrr,rrvn,mpr}.

Nevertheless, in our opinion the formulas $\left( \text{\ref{mf}}\right)
,\left( \text{\ref{mf1}}\right) \;$could give a simple pure geometric set up
in order to improve our knowledge about three dimensional gauge theories.
This is our motivation for the present letter.

The paper is organized as follows. Sect.2 is devoted to the computation of
the coefficients $\vartheta _\mu ^n(D,F)$ up to the fourth order in the $1/m$
expansion. In Sect.3 we present a simple classical cohomological argument
which supports the formulas $\left( \text{\ref{mf}}\right) ,\left( \ref{mf1}%
\right) $. Sect.4 deals with the $N=1$ superspace generalization. Finally,
we conclude with a few remarks concerning possible further applications.

\section{Some computations}

In order to have a more precise idea of the coefficients $\vartheta _\mu
^n(D,F)\;$let us give here the explicit value of some of them. Their
computation is really straightforward, one only needs to insert the eq.$%
\left( \text{\ref{mf1}}\right) \;$into the eq.$\left( \text{\ref{mf}}\right)
\;$and identify the terms with the same power in $1/m$. For instance, the
first four coefficients are found to be

\begin{eqnarray}
\vartheta _\mu ^1 &=&\frac 14\varepsilon _{\mu \sigma \tau }F^{\sigma \tau
}\;,  \label{coeff} \\
\vartheta _\mu ^2 &=&\frac 18D^\sigma F_{\sigma \mu }\;,  \nonumber \\
\vartheta _\mu ^3 &=&-\frac 1{16}\varepsilon _{\mu \sigma \tau }D^\sigma
D_\rho F^{\rho \tau }\;+\frac g{48}\varepsilon _{\mu \sigma \tau }\left[
F^{\sigma \rho },F_\rho ^{\;\tau }\right] \;,  \nonumber \\
\vartheta _\mu ^4 &=&-\frac 5{128}D^2D^\rho F_{\rho \mu }+\frac 5{128}D^\nu
D_\mu D^\lambda F_{\lambda \nu }\;  \nonumber \\
&&-\frac 7{192}g\left[ D^\rho F_{\rho \tau },F_\mu ^{\;\;\tau }\right]
\;-\frac 1{48}g\left[ D_\nu F_{\mu \lambda },F^{\lambda \nu \;}\right] \;. 
\nonumber
\end{eqnarray}
Observe that, as already remarked, all the coefficients of eq.$\left( \text{%
\ref{coeff}}\right) \;$are covariant, depending only on $F_{\mu \nu }\;$and
its covariant derivatives. Although the higher order coefficients can be
easily obtained, let us now focus on a cohomological argument which will
justify the formulas $\left( \text{\ref{mf}}\right) ,\left( \text{\ref{mf1}}%
\right) $.

\section{A cohomological argument}

In order to provide a cohomological argument for the eqs.$\left( \text{\ref
{mf}}\right) ,\left( \text{\ref{mf1}}\right) $ we shall make use of the
BRST\ antifield formulation. The construction of the corresponding classical
Slavnov-Taylor (or master equation) identity does not present any difficulty
and is easily carried out \cite{gmrr}. Only two antifields\footnote{%
As it is well known \cite{ht} the introduction of the antifields allows to
implement in cohomology the classical equations of motion.} $(A_\mu
^{*},c^{*})$ are needed, corresponding respectively to the gauge connection $%
A_\mu \;$and to the Faddeev-Popov ghost $c$. Let us observe now that within
the BRST framework the reabsorption of the pure Yang-Mills term $\int FF\;$%
through a gauge field redefintion, as it is implied by the eqs.$\left( \text{%
\ref{mf}}\right) ,\left( \text{\ref{mf1}}\right) $, lies in the possibility
of (re)expressing $\int FF\;$in the form of an exact BRST cocycle. That this
is indeed the case follows from a simple inspection of the BRST
transformation of the antifield $A_\mu ^{*}$, \textit{i.e. }$\;$

\begin{equation}
sA_\mu ^{*}=\frac 12\varepsilon _{\mu \nu \rho }F^{\nu \rho }+\frac 1mD^\nu
F_{\mu \nu }+\left\{ c,A_\mu ^{*}\right\} \;,  \label{sas}
\end{equation}
$s$ denoting the BRST differential.

The last term in eq.$\left( \text{\ref{sas}}\right) \;$states the simple
fact that under a rigid gauge transformation the antifield $A_\mu ^{*}\;$%
transforms according to the adjoint representation of the gauge group, and
can be neglected when the BRST differential $s$ acts on the space of the
gauge invariant quantities, as for instance $\int FF$. Contracting\footnote{%
We use here the euclidean normalization $\varepsilon^{\mu\nu\rho}%
\varepsilon_{\mu\sigma\tau}=(\delta^{\nu}_{\sigma}\delta^{\rho}_{\tau}
-\delta^{\rho}_{\sigma}\delta^{\nu}_{\tau})$.} now both sides of eq.$\left( 
\text{\ref{sas}}\right) \;$with $\varepsilon _{\mu \nu \rho },\;$the eq.$%
\left( \text{\ref{sas}}\right) $ can be cast in the following more
convenient form

\begin{equation}
F_{\mu \nu }=s(\varepsilon _{\mu \nu \rho }A^{*\rho })-\frac 1m\varepsilon
_{\mu \nu \rho }D_\lambda F^{\rho \lambda }\;-\left\{ c,\varepsilon _{\mu
\nu \rho }A^{*\rho }\right\} \;.  \label{ff}
\end{equation}
It becomes now apparent that the above formula\footnote{%
A similar cohomological argument has been already used by \cite{mpr} (see
for instance eqs.(6.31)) in the algebraic analysis of the $N=2$ version of
the topological massive Yang-Mills.} allows us to replace in any gauge
invariant quantity the field strength $F_{\mu \nu }\;$by a pure BRST
variation with, in addition, a term of the order $1/m$ containing a
covariant derivative $D_\mu $, \textit{i.e.}

\begin{eqnarray}
tr\int d^3xF_{\mu \nu }F^{\mu \nu } &=&tr\int d^3x\;\left( s(A^{*\mu
}\varepsilon _{\mu \nu \rho }F^{\nu \rho })\;-\frac 1mF^{\mu \nu
}\varepsilon _{\mu \nu \rho }D_\lambda F^{\rho \lambda }\right)  \nonumber
\label{rec-f} \\
&=&tr\int d^3x\;s\left( A^{*\mu }\varepsilon _{\mu \nu \rho }F^{\nu \rho
}\;+\frac 2mA^{*\mu }D^\sigma F_{\sigma \mu }\right) \;\;+\;O(\frac
1{m^2})\;\;  \nonumber \\
&=&.............\;\;\;.  \label{rec-f}
\end{eqnarray}
The expression $\left( \text{\ref{ff}}\right) \;$yields then a recursive
procedure since $F_{\mu \nu }\;$appears on both sides. At each step of the
iteration a new factor $(\varepsilon D/m)\;$will appear in the right hand
side of eq.$\left( \text{\ref{rec-f}}\right) $. We will end up therefore
with an infinite power series whose generic element of the order $n$ is
characterized by the presence of a factor of the kind $(\varepsilon D/m)^n$.
The eq.$\left( \text{\ref{ff}}\right) \;$implies thus that the Yang-Mills
term $\int FF\;$can always be rewritten as a pure BRST variation of an
infinite series in the expansion parameter $1/m$.\ It is this property which
allows us to reabsorb $\int FF$ into the Chern-Simons action by means of a
nonlinear gauge field redefinition. Moreover, all coefficients will be 
\textit{local }and \textit{covariant}\footnote{%
The covariance of the coefficients $\theta _\mu ^n$ in eq.$\left( \ref{mf1}%
\right) $ easily follows from the requirement of gauge invariance of $%
\mathcal{S}_{CS}(\widehat{A})$. As a consequence, the redefined field $%
\widehat{A}$ transforms as a connection under gauge transformations.}$.$ 
\textit{\ }

Let us conclude this section with the following remark. If we had started
with the pure Yang-Mills term as initial action, it would be impossible to
reach such a kind of conclusion. In fact the left hand side of the
expression $\left( \text{\ref{ff}}\right) \;$would be vanishing, due to the
absence of the Chern-Simons term. The formula $\left( \text{\ref{ff}}\right) 
$ would become thus useless. However, as soon as the Chern-Simons is
switched on, the Yang-Mills term can be seen as being generated by pure
Chern-Simons by means of a local and covariant gauge field redefinition.

\section{Supersymmetric generalization}

It is very easy to generalize the previous set up to the supersymmetric
version of topological massive Yang-Mills $\left( \text{\ref{tmym}}\right) $%
. Considering for instance the case of $N=1\;$in superspace we have,
following \cite{rrvn},

\begin{equation}
\mathcal{S}_{CS}(\Gamma )=-\frac 12tr\int dV\left( \Gamma ^\alpha D^\beta
D_\alpha \Gamma _\beta +\frac g3\Gamma ^\alpha \left[ \Gamma ^\beta
,D_{(\beta }\Gamma _{\alpha )}\right] +\frac{g^2}6\Gamma ^\alpha \left[
\Gamma ^\beta ,\left\{ \Gamma _\alpha ,\Gamma _\beta \right\} \right]
\right) \;,  \label{n1cs}
\end{equation}
and

\begin{equation}
\mathcal{S}_{YM}(\Gamma )=\frac 1mtr\int dV\;W^\alpha W_\alpha
\;,\;\;\;\;\;\;\;\;\;\;dV=d^3x\;d^2\theta \;,  \label{n1ym}
\end{equation}
where $\Gamma ^\alpha \;$is the spinor gauge superfield and $W_\alpha \;$is
the superfield strength given by

\begin{equation}
W_\alpha =D^\beta D_\alpha \Gamma _\beta +g\left[ \Gamma ^\beta ,D_\beta
\Gamma _\alpha \right] +\frac{g^2}3\left[ \Gamma ^\beta ,\left\{ \Gamma
_\alpha ,\Gamma _\beta \right\} \right] \;,  \label{n1sft}
\end{equation}
with $D_\alpha \;$being the ordinary superspace supersymmetric derivative ($%
\alpha ,\beta \;$are now spinor indices). Introducing the covariant
supersymmetric gauge derivative

\begin{equation}
\nabla _\alpha =D_\alpha +g\left[ \Gamma _\alpha ,\;\right] \;,  \label{scd}
\end{equation}
for the first coefficients $\vartheta _\alpha ^n(\nabla ,W)$ of the
supersymmetric version of the expansion $\left( \text{\ref{mf1}}\right) $ we
get

\begin{eqnarray}
\vartheta _\alpha ^1(\nabla ,W) &=&-W_\alpha \;,  \label{n1coeff} \\
\vartheta _\alpha ^2(\nabla ,W) &=&-\frac 12\nabla ^\beta \nabla _\alpha
W_\beta \;,  \nonumber \\
\vartheta _\alpha ^3(\nabla ,W) &=&-\frac 12\nabla ^\beta \nabla _\alpha
\nabla ^\gamma \nabla _\beta W_\gamma +\frac 13g\left[ W^\beta ,\nabla
_\alpha W_\beta \right] \;.  \nonumber
\end{eqnarray}
Again, they are all covariant. Of course, the previous cohomological
argument applies to the present supersymmetric case as well.

\section{Conclusion}

Let us conclude with a few comments and remarks on possible further
applications.

\vspace{5mm}

The first remark concerns the use of a nonlinear field redefinition.
Although the present work deals with pure classical considerations, it is
worthy to remind that nonlinear field redefinitions have already been used
in several cases, being typically needed when dimensionless fields are
present. This is the case, for instance, of the two dimensional nonlinar
sigma model \cite{bms} and of the $N=1$ superspace super Yang-Mills theories 
\cite{ps}. Let us underline that these nonlinear field redefinitions are, in
analogy with our case, completely\textit{\ local} and given explicitely by
an infinite power series in the fields. Other kinds of nonlinear but \textit{%
nonlocal } field redefinitions are also known. They are used in order to
compensate the nonlocal divergences which arise when noncovariant gauges are
employed. An example of such a kind of nonlinear nonlocal field redefinition
is provided by \cite{lm} who analysed in fact the one loop renormalization
of massive topological Yang-Mills theory in the light-cone gauge. It should
be clear, however, that this nonlinear nonlocal field redefinition is
completely different from that of eq.$\left( \text{\ref{mf1}}\right) .$

\vspace{5mm}

Although the higher order coefficients $\vartheta _\mu ^n\;$of the series $%
\left( \text{\ref{mf1}}\right) \;$can be computed straightforwardly, the
possibility of obtaining a closed expression for the infinite expansion $%
\left( \text{\ref{mf1}}\right) $ is very tempting. We limit to observe here
that the covariant character of the coefficients $\vartheta _\mu ^n$
naturally remind us the normal coordinate expansion frequently used in the
nonlinear two dimensional sigma models \cite{bms}.

\vspace{5mm}

Let us point out, finally, that the recursive cohomological argument of eq.$%
\left( \text{\ref{ff}}\right) \;$applies also in the case in which terms of
higher order in $F$ and its covariant derivatives $(FD^2F,...,etc.)$ are
introduced in the classical initial action. As one can easily understand,
the recursive cohomological argument relies essentially only on the presence
of the Chern-Simons term. Of course, this follows from the fact that the
variation of the Chern-Simons term yields exactly the field strength $F$. It
is this basic property which is at the origin of the recursive cohomological
argument $\left( \text{\ref{ff}}\right) .\;$In other words, provided the
Chern-Simons term is present in the game, the recursive argument still holds
for whatever metric dependent gauge invariant Yang-Mills type term one likes
to start with.

\vspace{5mm}

Possible quantum aspects related to the nonlinear field redefinition $\left( 
\text{\ref{mf1}}\right) $ as well as to a pure algebraic analysis of the
ultraviolet finiteness of topological massive Yang-Mills are under
investigation.

\vspace{5mm}

{\Large \textbf{Acknowledgements}}

We are grateful to Jos\'{e} Abdalla Helay\"{e}l-Neto and Gentil Oliveira
Pires for useful discussions and remarks. The Conselho\ Nacional de Pesquisa
e Desenvolvimento, CNP$q$ Brazil, the Faperj, Funda\c {c}\~{a}o de Amparo
\`{a} Pesquisa do Estado do Rio de Janeiro and the SR2-UERJ are gratefully
acknowledged for financial support.

\end{document}